\documentclass[11pt,twoside]{article}
\usepackage{asp2004}
\usepackage{psfig}
\usepackage{epsf}
\usepackage{graphics}
\usepackage{lscape}
\def\edcomment#1{\iffalse\marginpar{\raggedright\sl#1\/}\else\relax\fi}
\reversemarginpar

\begin{document}
\title{Implications of aging in young supernovae
}
\author{Poonam Chandra \& Alak Ray}
\affil{Tata Institute of Fundamental Research, Mumbai -400 005}


\begin{abstract}
We present here the combined radio spectrum from the
Giant Metrewave Radio Telescope (GMRT)
and Very Large Array (VLA) of  
type-IIb SN 1993J (age 11 years)
and a type-Ic  SN 2003bg (age 1 year). In SN 1993J, we 
find a break in the spectrum at 4 GHz and associate it with 
Synchrotron cooling break.  Hence, we
 determine the magnetic field
independent of equipartition assumption between relativistic
electrons and magnetic energy. We also see a hint of break 
in the spectrum of SN 2003bg between 22-40 GHz. The spectrum of SN 2003bg 
is well described by the synchrotron self absorption model.
\end{abstract}

\thispagestyle{plain}

\section{Introduction}

Synchrotron aging
can provide wealth of information 
related to plasma conditions in young supernovae.
We discuss two supernovae in this context
- an eleven years old type-IIb supernova SN 1993J and a one year
old type-Ic  SN 2003bg.  

\section{SN 1993J in M81}
We observed SN 1993J around day 3200 with the GMRT
in 610, 235 and 1420 MHz bands
combined this
dataset with the high frequency VLA observations
provided by C. Stockdale \& collaboration
(see \citet{cha04}).
The spectrum suggests a break in the
spectral index ($\Delta \alpha= 0.6$) in the optically 
thin part of the spectrum 
at 4 GHz (Fig. 1).
This variation is consistent with that
predicted from the synchrotron cooling effect
with continuous injection \citep{kar62}.
Including the effect of acceleration and adiabatic loss processes,
and using size of the SN $R=2.65\times 10^{17}$ cm on day 3200
from VLBI \citep{bar02}, we obtain magnetic field
$B=0.33\pm0.01 $ G \citep{cha04}.
From the best fit in SSA, the magnetic
field under equipartition assumption is $B_{eq}=38\pm17$ mG.
Comparison of the two magnetic field determines the
the ratio of relativistic energy of
particles to magnetic field energy, 
which is $ (0.85-40) \times 10^{-5}$.

\begin{figure}
\plottwo{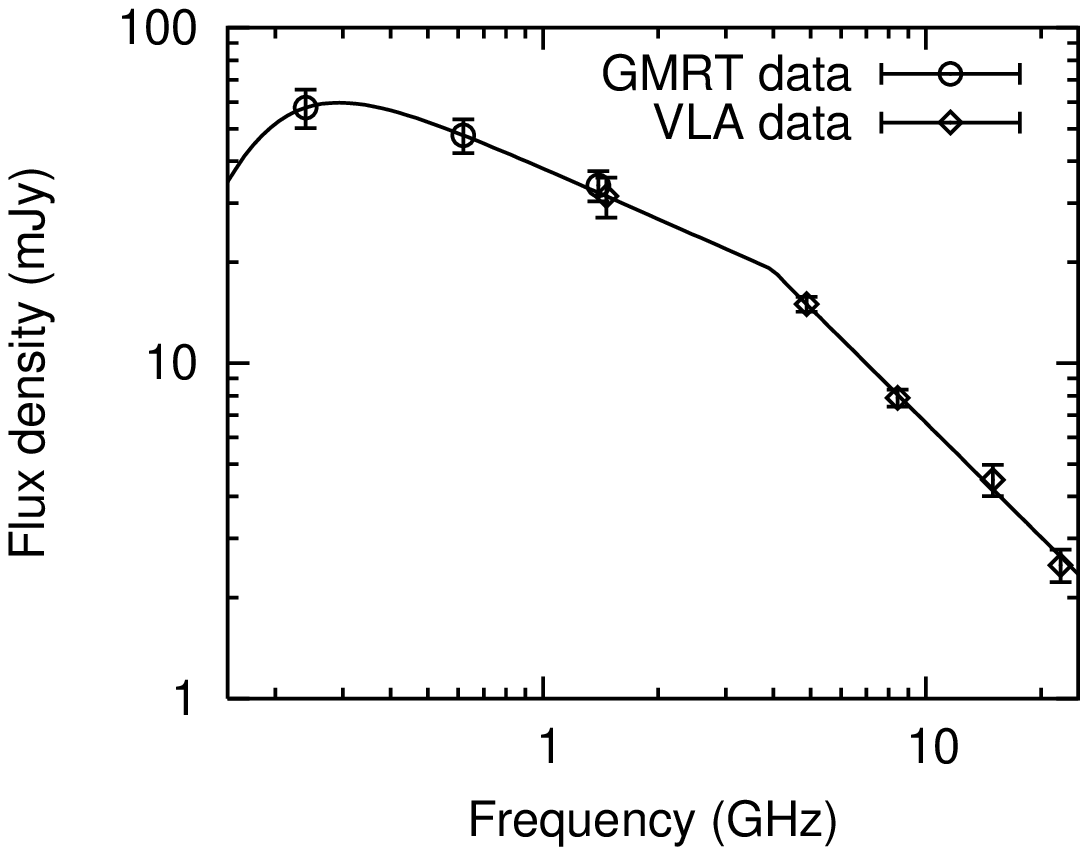}{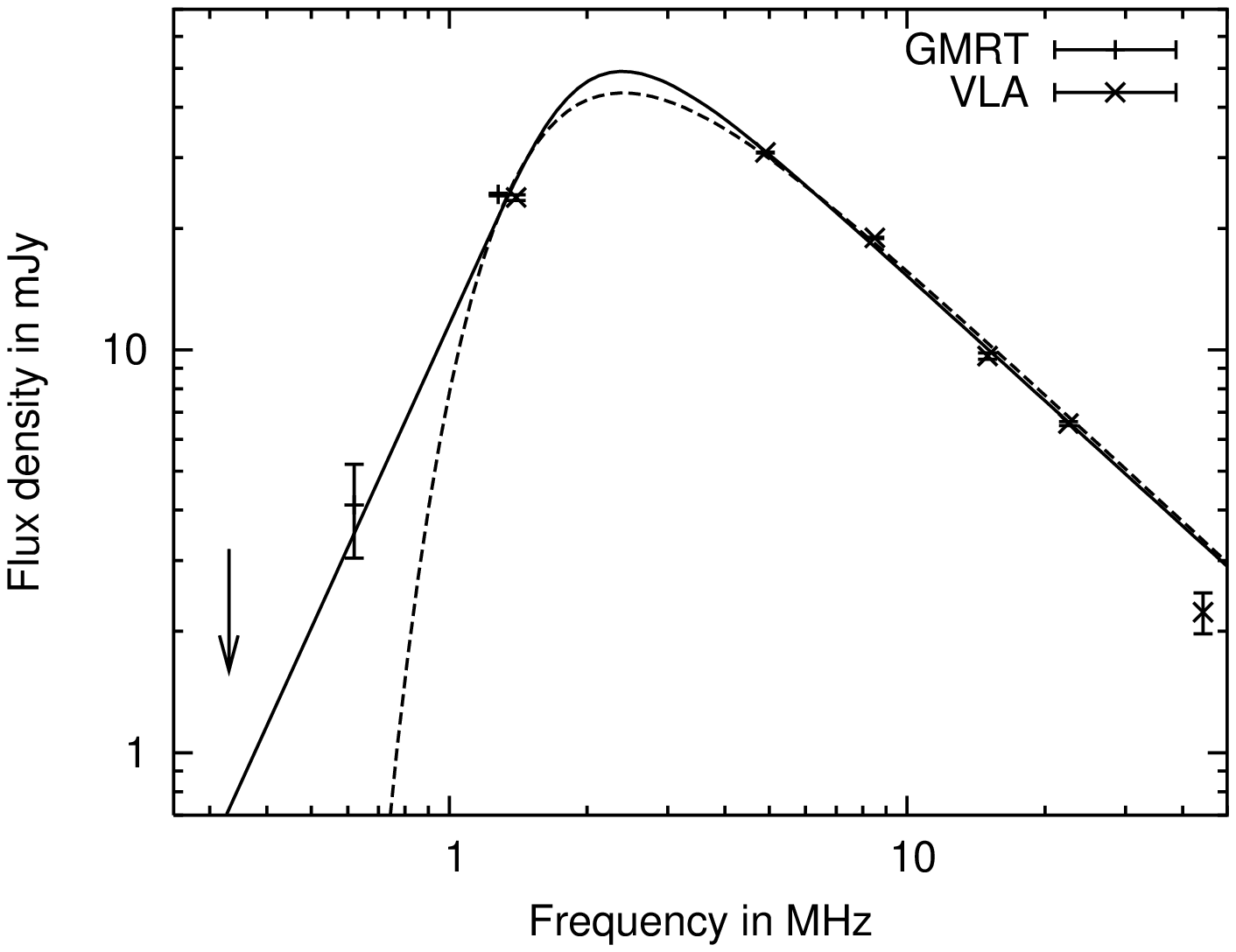}
\caption{Left Panel: Combined GMRT and VLA spectrum of SN 1993J
(day 3200).
Solid line shows the SSA fit to the data with
the break in the spectrum ($\Delta \alpha=0.6$) 
at 4 GHz. Right panel: 
SSA (solid line) and FFA (dashed line) fits to
the combined GMRT and VLA spectrum of SN 2003bg.}
\end{figure}

\section{SN 2003bg in MCG -05-10-015}
                                                                                
SN 2003bg is type Ic supernova in MCG -05-10-015 (19 Mpc). It was discovered
on 2003 Feb 25, most likely two weeks after the explosion.
We observed SN 2003bg with GMRT in 1280, 610 and 325 MHz
bands between 2003 Feb 2- Feb 8
and it was observed at high VLA frequencies on 2003 Feb 8
by A. Soderberg and S. Kulkarni, thus obtaining
the radio spectrum covering $0.3-44.0$ GHz.
We fit  homogeneous
Free-free absorption (FFA) and Synchrotron Self Absorption (SSA) models 
to the spectrum (see Fig. 1).
The 610 MHz data point
 clearly eliminates the FFA model and favors the SSA model.
We find the  following parameters from our fits:  spectral
index ($\alpha = 3$), size of SN 2003bg ($R= (9.98 \pm 0.43) \times
10^{16}$ cm), expansion speed ($ v = 33,000\,{\rm km \, s^{-1}}$),
and magnetic field ($ B = 0.18 \pm 0.03$ G).
There seems to be hint of a break around 30 GHz, although this
could be due to calibration problems at high frequencies. 
We need the radio spectrum 
extended beyond 40 GHz to determine if the break is real.
We are planning simultaneous observations of SN 2003bg with GMRT along with VLA
and ATCA. This will provide us the spectrum from 0.2 GHz to 80 GHz
and will establish whether the spectral break is real or due to data artefact.

\section{Discussion and conclusion}

The relativistic particle energy density is 
far below than the magnetic energy density
(by a factor of 1/10000).
This is quite unlike the expectation of efficient shock acceleration
of relativistic particles in SNe. The expected fraction of post-shock 
pressure in relativistic particles is $W \ge 0.1$ as 10\% of the energy ejected
by the SNe is needed to power the cosmic rays in the galaxy \citep{mck87}.
It indicates that perhaps the efficiency of acceleration is either highly 
variable in SNe, or it will evolve with time in SN 1993J.

\acknowledgments{
We thank K. Weiler \&
collaboration, and 
A. Soderberg \& S. Kulkarni for data on SN 1993J and
2003bg respectively.
We thank GMRT staff for making the observations possible.
}

\end{document}